\documentstyle[11pt]{article}
\newcommand{\blankline}{\vskip .3cm}
\newcommand{\f}{\begin{equation}}
\newcommand{\ff}{\end{equation}}

\begin{document}
\centerline{\LARGE Using neutron stars and primordial black holes}
\blankline
\rm
\centerline{\LARGE to test theories of quantum gravity}
\centerline{Lee Smolin${}^*$}
\blankline
\centerline{\it  Center for Gravitational Physics and Geometry}
\centerline{\it Department of Physics}
 \centerline {\it The Pennsylvania State University}
\centerline{\it University Park, PA, USA 16802}
 \blankline
 \blankline
 \blankline
\centerline{October 31, 1997}
 \blankline
 \blankline
 \blankline
\centerline{ABSTRACT}
Three observational tests of cosmological natural
selection, a  theory that follows from some hypotheses about
quantum gravity, are described.  If true, this theory explains
the choices of the parameters of the standard model of
particle physics.   The first,
the observation of a pulsar with mass greater than $2.5 M_\circ$,
would cleanly refute the theory.  The second and third, having to
do with primordial black holes and early massive star formation,
could do so given likely developments in the near future.  
However given present knowledge these arguments do not
presently refute the theory.  This
shows that cosmological 
natural selection has not so far been refuted,while
remaining very
vulnerable to falsification by possible observations.

\blankline
${}^*$ smolin@phys.psu.edu
\eject

\section{Introduction}

It is unfortunate that the Planck scale and unification
scale are so remote from what can be probed experimentally
that 
most hypotheses about quantum gravity and the unification
of the different interactions are developed without benefit of
experimental test.  In order to counteract this, 
one may try to adopt a strategy of searching for
hypotheses about fundamental physics whose main merit is that
they are testable given present knowledge.  One way to do this
which has often been pointed out is to use the apparent fact that
very high temperatures, densities and energies were experienced,
however briefly, in the early universe.  Among other things, this
has led to the hypothesis of inflation\cite{newinflation,tiltinflation},
certain versions of
which are going to be well tested in forthcoming observations of
the MAP and PLANCK missions.

In \cite{evolution,moreev,lotc} a theory aimed at explaining the 
parameters of the
standard model of particle physics was introduced, which 
assumes the following two hypothesis about fundamental
physics: 

\begin{enumerate} 

\item{}Black hole singularities bounce, leading to new 
expanding regions of spacetime, one per each black hole.  

\item{}When this happens the dimensionless 
parameters of the standard model of 
low energy physics of the new region differ by a {\it small }
random change from those in the region in which the black hole
formed.   Small here means with small with 
respect to the change that would be required to significantly change
$B(p)$, the expected number of black holes produced in the
classical region of spacetime produced by the bounce.  
Here $p \in {\cal P}$, the space of dimensionless parameters
of the standard model.   

\end{enumerate}

With the exception of the {\it small} in 2), these are not
new hypotheses, and have been discussed, for example in 
\cite{old}.
Their conjunction leads to a predictive theory, because, using
standard arguments from population biology, after many
iterations from a large set of random starts, the population
of regions, given by a distribution $\rho (p)$, is peaked around
local extrema of $B(p)$.  With more detailed assumptions more
can be deduced, but this is sufficient to lead to observational
tests of hypothesis 1) and 2) because this implies the
statement that: 
 
\begin{itemize}
\item{}$\cal S$:
{\it If $p$ is changed from the present value in any direction in
$\cal P$ the first significant changes in $B(p)$ encountered
must be to decrease $B(p)$.}   

\end{itemize}

The conjunction of 1) and 2)
thus constitute a theory that if true would explain the values
of the parameters of the standard model without recourse to
the Anthropic Principle.  This theory may be called, ``cosmological
natural selection."  It should be emphasized that it is completely
consistent with our knowledge of fundamental 
physics\footnote{Other approaches to cosmology which employ
phenomena analogous to biological
evolution have been proposed, including
Davies\cite{davies}, Gribbin\cite{gribbin}, Kauffman\cite{stu},
Linde\cite{andrei}, Nambu\cite{nambu},Schweber, 
Thirring and Wheeler\cite{wheeler}.  
The best developed of these is a series of papers of Linde and collaborators
\cite{moreandrei} in the context of inflationary cosmology.}.For
example, recent work in string theory has revealed that that theory
has a large number of stable vacua, or phases, in which the
standard model parameters differ.  
When string theory
becomes better understood present knowledge seems 
to indicate that the likely effect will be not to fix
$p$ in $\cal P$ but to replace it with a microscopic parameter
$m$ in the space $\cal M$ of string vacua.  It should also be
mentioned that the possibility of a bounce has been discussed
in several different approaches to quantum gravity, including
string theory\cite{bounce}.

Several arguments were made that $\cal S$ is in
fact contradicted by present observation \cite{ellis,harrison,silk}.  
These were found to
depend either on confusions about the hypothesis itself or 
on too simple assumptions about star formation and
are thus invalid\footnote{See especially the appendix
of \cite{lotc}, which addresses most of the objections 
published in \cite{ellis,silk} and elsewhere.}.   Other claims 
have been made
that with present knowledge $\cal S$ is in fact not 
testable\cite{silk,rees}.  Here I would like to show that these claims
are also false, by  explaining why a single observation of an
astrophysical object that very well might exist-a heavy neutron
star-
would refute
$\cal S$.    After this I describe two more kinds of observations
that may be made in the near future which could lead to 
refutations of $\cal S$.  These have to do with more accurate
observations of  the
spectrum of fluctuations in the cosmic microwave background
($CMB$) and the initial mass function for star formation in
the absence of carbon.  

\section{Why a single heavy pulsar would refute $\cal S$.}

Bethe and Brown, in \cite{bethebrown} introduced 
the hypothesis that neutron
star cores contain a condensate of $K^-$ mesons.  For the present
purposes their work can be expressed in the following way.
Calculations show\cite{bethebrown} that there is a 
critical value $\mu_c$
for the strange quark mass $\mu$ such that if $\mu < \mu_c$ then
neutron star cores consist of approximately equal numbers of
protons and neutrons with the charge balanced by a condensate
of $K^-$ mesons.  The reason is that in nuclear matter the effective
mass of the $K^-$ is renormalized downward by an
amount depending on the density $\rho$.   Given a choice of
the strange quark mass,
$\mu$, let  $\rho_0 (\mu )$ be the density 
where the renormalized Kaon mass is less than the electron mass.
$\mu_c$ is the value of $\mu$ where $\rho_0 (\mu )$ is less
than the density $\rho_e$ at which the electrons react with the
protons to form neutrons.  In either case one neutrino per 
electron is produced, leading to a supernova.  

Bethe, Brown and collaborators claim that calculations show
that $\mu <\mu_c$ \cite{bethebrown}.  But their calculations involve 
approximations
such as chiral dynamics and may be sufficiently inaccurate
that in fact $\mu_c > \mu$.  However, the
accuracy of the calculations 
increases as $\mu^{-2}$ as $\mu$ is decreased so, even
if we are not sure of the conclusion that $\mu < \mu_c$, we
can be reasonably sure of the existence of such a critical value
$\mu_c$.  Then we may reason as follows.  
If $\mu < \mu_c$ then, as shown by calculations\cite{bethebrown}
the upper
mass limit is low, approximately $1.5 M_\circ$.  If $\mu> \mu_c$
neutron stars have the conventional equations of state and the
upper mass limit is higher, almost certainly above $2$
\cite{uml}.  
Therefor a single observation of a neutron star whose mass
$M$ was sufficiently high would show that $\mu > \mu_c$,  
refuting Bethe and Brown's claim for the opposite.  
Sufficiently high is certainly $2.5 M_\circ$, although if
one is completely confident of Bethe and Brown's upper limit
of $1.5$ solar masses, any value higher than this would be
troubling.
Furthermore,
this would refute $\cal S$ because it would then be the case that
a decrease of $\mu$ would lead to a world with a lower upper
mass limit for neutron stars, and therefor more black holes.

Presently all well measured neutron star masses are from
binary pulsar data and are all below 
$1.5 M_\circ$ \cite{nsm}\footnote{Other methods yield less
precise estimates\cite{estimates}.}. But an 
observation of a heavy neutron star may be made at any time.  

We may note that this argument is independent of any issue
of selection effects associated with ``anthropic reasoning",
because the value of the strange quark mass $\mu$ may
be varied within a large range before it produces a significant
effect on the chemistry\footnote{Skeptics might reply that
were $\cal S$ so refuted it could be modified to a new
${\cal S}^\prime$,  which was not refuted by the addition of the
hypothesis that $\mu$ is not an independent parameter and cannot
be varied without also, say, changing the proton-neutron mass
difference, leading to large effects in star formation.  It is of course,
a standard observation of philosophers of science that most 
scientific hypotheses can be saved from refutation by the 
proliferation
of ad hoc hypotheses.  In spite of this, science proceeds by
rejecting hypotheses that are refuted in the absence of special
fixes.  There are occasions where such a fix is warranted.  The 
present case  would only be among them if there were a prefered
fundamental theory, such as string theory, which had strong
independent experimental support, in which it turned
out that $\mu$ was in fact not an independent parameter, but
could not be changed without altering the values of parameters
that strongly affect star formation and evolution. }.

\section{How observations of the $CMB$ could refute $\cal S$.}

It may be observed that the universe might have had many more
primordial black holes than it seems to have were the
spectrum of primordial fluctuations, $f(n)$ tilted 
to increase the
proportion on small scales\cite{silk}.  Of course, this observation by
itself does not refute $\cal S$ directly unless it is shown that
the standard model has a parameter that can be varied to
achieve the tilt in the spectrum.  It does not, but it is reasonable
to examine whether plausible extensions of the standard model
might.  One such plausible extension is to add fields that could
serve as an inflaton, so that the theory predicts inflation.  Given
an extention of the standard model, $\cal E$,  
that predicts inflation, the
spectrum of primordial fluctuations is in fact predicted as a 
function of the parameters of $\cal E$.  Thus, $\cal S$ is
refuted if a) some model $\cal E$ of inflation is observationally
confirmed and if b) that particular extension of the standard
model has some parameter, $p_{inf}$ that can be modified to
{\it increase the total numbers of black holes produced, including
primordial black holes.}  Given
the accuracy expected for observations of the $CMB$ from 
MAP and PLANCK, there is a realistic possibility that observations
will distinguish between different hypotheses for $\cal E$ and measure
the values of their parameters.

In the standard ``new" inflationary scenario\cite{newinflation} 
there is no
parameter that fulfills the function required of $p_{inf}$.   There
is the inflaton coupling $\lambda$, and it is true that the
amplitude of the $f(n)$ is proportional to $\lambda$ so that
the number of primordial black holes can be increased by
increasing $\lambda$.  However, the size of the region that
inflates $R$, is given by $R \approx e^{\lambda^{-1/2}}$.  This
effect overwealms the possibility of making primordial black holes.
In fact, if the observations confirm that the standard new inflationary
scenario is correct, then $\cal S$ is refuted if $\lambda$ is not
tuned to the value that results in the largest total
production of black holes in the inflated region\cite{evolution}.
Because of the
exponential decrease in $R$ with increasing $\lambda$, this
is likely close to the {\it smallest} possible value that leads to a 
sizable constant density of black holes produced in comoving
volumes during the history of the universe.  This is likely
the smallest $\lambda$ that still allows prolific formation of
galaxies\cite{evolution}.

This seems consistent
with the actual situation in which there appears to have been
little production of primordial black holes, so that the primary mode
of production of black holes seems to be through massive star 
production, in galaxies that apparently do not form till rather late,
given that $\delta \rho /\rho \approx 10^{-5}$.   Of course, the
observations that favor $\Omega \approx .3$ are also troubling
for the standard new inflation which predicts 
$\Omega$ very near one.

However, there are non-standard models of inflation that have
parameters $p_{inf}$ that can be varied from the present
values in a manner that tilts $f(n)$ so that more primordial
black holes are created than in our universe, without at the same
time decreasing $R$\cite{tiltinflation}.  If future observations 
of the $CMB$ cleanly
show that the standard new inflation is ruled out, and only models
with such a parameter $p_{inf}$ are allowed, then 
$\cal S$ willl be refuted.  

This is a weaker argument than the first one,  
but given the scope for increases in
the accuracy of measurements of the $CMB$, and hence of
tests of inflationary models, it is a realistic possibility that
$\cal S$ may be refuted by such an argument.

\section{How early star formation could refute $\cal S$.}

As shown in \cite{carrees,rees,evolution} there 
are several directions in $\cal P$
which lead to universes that contain no stable nuclear
bound states.  It is argued  in \cite{evolution,lotc} this 
leads to a strong
decrease in $B(p)$ because the gravitational collapse of objects
more massive than the upper mass limit of neutron stars
in our universe seems to depend on the cooling mechanisms
in giant molecular clouds, which are dominated by 
radiation from $CO$.  In a universe without nuclear bound
states the upper mass limit for stable collapsed objects is
unlikely to decrease dramatically (as the dominant factor
ensuring stability is fermi statistics)  while without cooling
from $CO$ collapsed objects whose ultimate size is above
the upper mass limit are likely to be less common.  

In the
absence of bound states the main cooling mechanism appears
to involve molecular hydrogen\cite{early}, but 
there are two reasons to
suppose this would not lead to plentiful collapse of massive
objects in a world with nuclear bound states.  The first is that
in such a world there would be no dust grains which appear to
be the primary catalysts for the binding of molecular hydrogen.
The second is that in any case molecular hydrogen is a  
less efficient coolant than $CO$.  

This is also a weaker argument than the first, given present
uncertainties in our understanding of star formation processes, but
as that understanding is likely to become more precise in the
near future let us follow it.  Could this argument 
be refuted by any possible
observations?  In the present universe the collapse of massive
objects is dominated by processes that involve nuclear bound
states, but we have available a laboratory for the collapse of
objects in the absence of nuclear bound states, which is the
universe before enrichment with metals.  Indeed, we know
that there must have been collapse of massive objects during
that period as otherwise carbon, oxygen and other elements
would not have been produced in the first place.   But given
that $CO$ acts as a catalyst for formation of heavy elements,
and that the dust formed from heavy elements produced in stars
is also a catalyst for molecular binding, 
there is an instability whereas any chance formation of massive
objects leads in a few million years to both an enrichment of the
surrounding medium and the production of significant quantities
of dust, and these greatly increase the probability for
the formation of additional massive objects.  Hence, the initial
rate of formation of heavy objects in the absence of enrichment
{\it does not have to be very high} to explain how the universe
first became enriched.

This shows  that the fact that there was
some collapse of heavy objects before enrichment does not refute
the argument that the number of black holes produced in a universe
without nuclear bound states would be much less than at
present.  But while it thus doesn't refute $\cal S$, it doesn't
establish it either.  It is still consistent with present knowledge
that the production of massive objects in the absense of
heavy elements proceeds efficiently under the right conditions,
so that there may have in fact been a great deal of early star
formation uncatalyzed by any process involving heavy elements.

This could lead to a refutation of $\cal S$ because, in a world
without nuclear bound states, many more  massive collapsed
objects would become black holes than do in our universe, where the
collapse is delayed by stellar nucleosythesis.  

The question is then 
whether a combination of observation and theory
could disentangle the strong catalytic effects of heavy elements
leading to a strong positive feedback in massive star formation
from the initial rate of massive star formation without heavy
elements.  
Although models of star formation with and without heavy
elements are not sufficiently developed to distinguish the
two contributions to early star formation, it is 
likey that this will become possible  as our ability to model star
formation improves.
If so than it is also possible that future observations will be
able to measure enough information about early star formation
to distinguish the two effects.  If the conclusion is that the number of
black holes formed is greater in world without nuclear bound
states than in our own then $\cal S$ would be refuted.

\section{Conclusions}

The first argument shows that a very likely astronomical
observation, the discovery of a pulsar with mass above
$2.5 M_\circ$ would refute $\cal S$.  This shows that the
theory of cosmological natural selection is falsifiable at present.
The other two arguments each lead to a case in which 
combinations of theoretical and observational developments
that are likely within the next decade could also easily
refute $\cal S$.  This shows that the vulnerability of $\cal S$
to falsification is very likely to strongly increase in the near future.

If in fact $\cal S$ is not refuted after many more pulsar 
masses are well measured and the links between theory and
observation in inflation and star formation are much improved, 
and if in the meantime 
string theory continues to be consistent with a large
choice of models and parameters of low energy physics, it 
will be difficult to aviod taking  cosmological natural selection
seriously as an explanation of the values of the standard model
parameters.   Given this, and the fact that even presently it seems
to be the most vulnerable to falsification of all the proposals so
far made to explain the parameters of the standard model, its
predictions deserve further study.  

\section*{ACKNOWLEDGEMENTS}

I am grateful to Martin Rees and Joseph Silk for correspondence on 
these questions.  Conversations and correspondence with
Jane Charlton,  Louis Crane,   
George Ellis, Bruce Elmegreen,  Eric Feigelson, Piet Hut, Anna Jangren, 
Stuart Kauffman and Peter Meszeros, 
have also been
very helpful in formulating tests of cosmological natural selection.
This work was supported by
NSF grant PHY-9514240 to The Pennsylvania State
University.

\end{document}